\documentclass{article}

\usepackage[english]{babel}

\usepackage[a4paper,top=2cm,bottom=2cm,left=3cm,right=3cm,marginparwidth=1.75cm]{geometry}

\usepackage{amssymb}
\usepackage{siunitx}
\PassOptionsToPackage{hyphens}{url}\usepackage{hyperref}
\usepackage{cleveref}
\usepackage[utf8]{inputenc}
\usepackage[right]{lineno}
\usepackage{csquotes}
\usepackage{booktabs}
\usepackage{longtable}
\usepackage{adjustbox}
\usepackage{array}
\usepackage{url}
\usepackage{titlesec}
\usepackage{authblk}
\usepackage{xcolor} % Load the xcolor package for color options

\titleformat{\subsection}
  {\mdseries\itshape\large} % Medium series, italic shape, and large font size
  {\thesubsection}{1em}{} % Numbering, spacing, and the section title itself

\usepackage[english]{babel}
\usepackage[style=authoryear,backend=biber,natbib=true,maxcitenames=2,uniquelist=false]{biblatex}
\DeclareNameAlias{sortname}{family-given}
\DeclareNameAlias{default}{family-given}

\renewbibmacro{in:}{}
\DeclareFieldFormat[article]{title}{\mkbibquote{#1}\addcomma}
\DeclareFieldFormat[book]{title}{\mkbibemph{#1}\addcomma}
\DeclareFieldFormat[bookinbook]{title}{\mkbibemph{#1}\addcomma}
\DeclareFieldFormat[inbook]{title}{\mkbibquote{#1}\addcomma}
\DeclareFieldFormat[incollection]{title}{\mkbibquote{#1}\addcomma}
\DeclareFieldFormat[inproceedings]{title}{\mkbibquote{#1}\addcomma}
\DeclareFieldFormat[manual]{title}{\mkbibemph{#1}\addcomma}
\DeclareFieldFormat[misc]{title}{\mkbibemph{#1}\addcomma}
\DeclareFieldFormat[thesis]{title}{\mkbibemph{#1}\addcomma}
\DeclareFieldFormat[unpublished]{title}{\mkbibquote{#1}\addcomma}
\DeclareFieldFormat[patent]{title}{\mkbibemph{#1}\addcomma}
\DeclareFieldFormat[report]{title}{\mkbibemph{#1}\addcomma}
\DeclareFieldFormat[online]{title}{\mkbibquote{#1}\addcomma}
\DeclareFieldFormat[software]{title}{\mkbibemph{#1}\addcomma}
\DeclareFieldFormat[booklet]{title}{\mkbibemph{#1}\addcomma}
\DeclareFieldFormat[periodical]{title}{\mkbibemph{#1}\addcomma}
\DeclareFieldFormat[standard]{title}{\mkbibemph{#1}\addcomma}

\DeclareFieldFormat[article]{journaltitle}{\iffieldundef{shortjournal}{\mkbibemph{#1}\addcomma}{\mkbibemph{\printfield{shortjournal}}\addcomma}}
\DeclareFieldFormat{volume}{\bibstring{volume}~#1}
\DeclareFieldFormat{number}{\bibstring{number}~#1}

\DefineBibliographyStrings{english}{
  volume = {Vol.},
  number = {No.}
}

\renewbibmacro*{volume+number+eid}{%
  \printfield{volume}%
  \setunit*{\addspace}%
  \printfield{number}%
  \setunit{\addcomma\space}%
  \printfield{eid}}

\renewbibmacro*{journal+issuetitle}{%
  \usebibmacro{journal}%
  \setunit*{\addcomma\space}%
  \usebibmacro{volume+number+eid}%
  \setunit{\addcomma\space}%
  \usebibmacro{issue+date}}

\renewbibmacro*{publisher+location+date}{%
  \printlist{publisher}%
  \iflistundef{location}
    {\setunit*{\addcomma\space}}
    {\setunit*{\addcolon\space}}%
  \printlist{location}%
  \setunit*{\addcomma\space}%
  \usebibmacro{date}}

\DeclareCiteCommand{\cite}[\mkbibparens]
  {\usebibmacro{prenote}}
  {\usebibmacro{citeindex}%
   \usebibmacro{cite}}
  {\multicitedelim}
  {\usebibmacro{postnote}}

\renewbibmacro*{cite:labelyear+extrayear}{%
  \iffieldundef{labelyear}
    {}
    {\printtext[bibhyperref]{%
       \printfield{labelyear}%
       \printfield{extrayear}}}}

\renewbibmacro*{cite:labeldate+extradate}{%
  \iffieldundef{labelyear}
    {}
    {\printtext[bibhyperref]{%
       \printfield{labelyear}%
       \printfield{extradate}}}}

\AtEveryBibitem{
  \clearfield{month}
  \clearfield{day}
  \ifentrytype{book}{
    \clearlist{location}
  }{}
}

\DefineBibliographyStrings{english}{
  andothers = {\textit{et al.},}
}

\DeclareFieldFormat[article]{volume}{\bibstring{jourvol}\addnbspace #1}
\DeclareFieldFormat[article]{number}{\bibstring{number}\addnbspace #1}
\DeclareFieldFormat[article]{volume}{Vol. #1}
\DeclareFieldFormat[article]{number}{No. #1}
\DeclareFieldFormat{url}{\bibstring{available at}\addcolon\space\url{#1}}
\DeclareFieldFormat{urldate}{\mkbibparens{accessed \addspace#1}}

\DeclareFieldFormat{urldate}{%
  \mkbibparens{accessed\space%
    \thefield{urlday}\addspace%
    \mkbibmonth{\thefield{urlmonth}}\addspace%
    \thefield{urlyear}}}

\crefformat{figure}{#2Figure~#1#3}
\Crefformat{figure}{#2Figure~#1#3}
\crefformat{table}{#2Table~#1#3}
\Crefformat{table}{#2Table~#1#3}
\crefformat{section}{#2Section~#1#3}
\Crefformat{section}{#2Section~#1#3}

\author[1]{Rishivarshil Nelakurti}
\author[2]{Christopher Hill}

\affil[1]{Olentangy High School, AP Research, Lewis Center, USA
(nelakurti.2@buckeyemail.osu.edu)}
\affil[2]{The Ohio State University, Physics, Columbus, USA}

\title{Evaluating Modifications to Classifiers for Identification of Higgs Bosons}

\begin{document}
\maketitle

\begin{abstract}
The Higgs boson, discovered back in 2012 through collision data at the Large Hadron Collider (LHC) by ATLAS and CMS experiments, marked a significant inflection point in High Energy Physics (HEP). Today, it's crucial to precisely measure Higgs production processes with LHC experiments in order to gain insights into the universe and find any invisible physics. To analyze the vast data that LHC experiments generate, classical machine learning has become an invaluable tool. However, classical classifiers often struggle with detecting higgs production processes, leading to incorrect labeling of Higgs Bosons. This paper aims to tackle this classification problem by investigating the use of quantum machine learning (QML).

%add 6 keywords
\end{abstract}

\section{High Energy Physics}
\label{sec:introduction}
\subsection{Overview of Higgs Boson}
The Higgs boson, which explains how particles gain mass, was first introduced in 1964. David J. Miller provides an analogy to understand the mechanism of particles gaining mass. He imagined a quiet room filled with physicists, representing the Higgs bosons. A famous scientist enters the room (a particle moves through the Higgs bosons), causes a disturbance, and attracts a group of admirers with every step. The gathering of admirers is a comparison to mass. The Higgs boson explains why certain particles have mass while others do not.

Moreover, the Higgs boson is the centerpiece of the standard model of particle physics, which serves as the current understanding of the actions of particles. The proof of the Higgs boson helps physicists to explain and investigate more of the standard model[11].

\subsection{Higgs Boson}
At the LHC, protons are accelerated at high speeds and collide together to produce new particles, such as the Higgs bosons. The discovery of the Higgs Boson was made possible by ATLAS and CMS, the two largest particle detectors designed to detect subatomic particles generated from these high-energy collisions. 

Since it cannot be directly observed, detecting the Higgs boson is a challenging task at the LHC. Additionally, the production of Higgs bosons is rare, with only one Higgs boson produced for every billion proton-proton collisions. This is because they are only produced through a specific type of interaction between the colliding protons. The LHC has to record only the most interesting collisions, which necessitates the use of algorithms that can handle big data classification[12]. 

Moreover, the Higgs boson decays rapidly into different particles, such as two photons, two taus, or two W bosons, making it difficult to detect directly. The decay occurs with very low probability, and various outcomes can occur in a single collision with different probabilities[13]. This study focuses solely on the Higgs to tau tau decay.

Another reason why the detection of the Higgs boson is challenging is because of the background noise generated by other particle interactions. It is difficult to differentiate the Higgs boson from this background noise[12].  Classical classifiers used to identify the signal and background noise often result in incorrect labeling of background processes, signals (label noise), and systematic errors[14]. To overcome this, researchers have looked at QML as its been theorized that quantum computers may outperform classical computers in machine learning tasks.

\section{Literature Review}
\label{sec:method}
\subsection{From Bits to Qubits: Historical Overview}
To understand quantum computers, one must understand how classical computers work. The basic operations in computers run off of a series of on and off, represented by a 0 or a 1, also called bits. Through the use of logic gates, by turning a 0 to a 1 and a 1 to a 0, different operations and their variations dictate the functions of the computer.

However, qubits have replaced classical bits as a result of the development of quantum computing, for two primary reasons: superposition and entanglement. According to John Preskill, while classical bits represent a 0 or a 1, qubits can exist as multiple states (superposition), which allows them to simultaneously represent a 0 and a 1. This enables calculations that are 2 to the power of n-1 times more complex[1]. This is because the operations will require far less qubits than it will bits, meaning the amount of bits is no longer a restrictive factor.

In the early days of computing, computers used vacuum tubes, then transistors, and then integrated circuits to perform calculations[2]. Smaller, quicker, and more effective computers were made possible by each of these technical advances. And as computing power increased, the limits of classical bits became more obvious.

The fact that classical bits can only represent a 0 or a 1 is one of their main drawbacks. In order to answer a problem, a classical computer needs to make several calculations, whereas a quantum computer can make the same calculation in only one.

Pairs of qubits can be entangled, which causes them to be linked. Entanglement is used as a multiplier for qubits. As one entangles more qubits, the ability of the system to make calculations exponentially grows rather than linearly.

\subsection{Quantum Algorithms}
To perform calculations on the qubits, operations called quantum gates are applied to qubits to manipulate their state[4]. Just like the logical gates of a classical computer, these quantum gates can, for instance, change the qubit from 0 to both 0 and 1. A combination of these gates can perform various complex computations and these combinations are called quantum circuits[4].

These circuits, when combined in specific ways, form quantum algorithms. Quantum gates and circuits are utilized to create algorithms that take full advantage of the peculiar attributes of qubits, like superposition and entanglement[5].

These algorithms are exclusively designed to run on a quantum computer and perform certain tasks much faster than classical algorithms, and are widely used today[4]. This paper aims to investigate the use of the three most popular classifiers in identifying the Higgs Boson: the Variational Quantum Classifier (VQC), Quantum Support Vector Machine (QSVM), and Quantum Neural Net (QNN). 

Both VQC and QSVM algorithms employ the technique of using a quantum circuit to map the input data into a higher-dimensional space, where the line between distinct data classes is more readily discernible[6]. 

The VQC algorithm is comprised of encoding input data and, unlike the other classifiers, trains variational circuits. The variational circuit is a parametrized quantum circuit that is trained to minimize the difference between the desired output and the actual output of the circuit[7]. The parameters of the variational circuit are adjusted iteratively to find the optimal solution.

VQC:
\begin{enumerate}
    \item Applies a feature map that feeds the classical data into the quantum circuit. In the feature map, the gates encode the input data to the quantum circuit by applying rotations of an angle \(\theta\) .
    \item Applies a parameterized circuit parameterized by gate angles \(\theta\) . The gate angles \(\theta\) are determined by the optimizer.
\end{enumerate}

\begin{figure}
    \centering
    \includegraphics[width=0.5\linewidth]{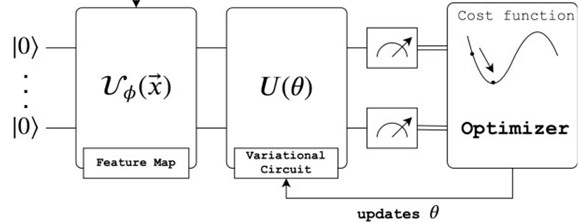}
    \caption{VQC Circuit}
    \label{fig:1}
\end{figure}

The next classifier, the QSVM, is based on the classical Support Vector Machine (SVM) algorithm, where like the classical SVM, QSVMs are based on the idea of finding the best boundary (hyperplane) that separates different classes of data. However, instead of using classical bits to represent the data, QSVMs use qubits and quantum gates to perform the calculations\textsuperscript{[6]}.

 The QSVM algorithm uses quantum gates to map the data into a high-dimensional space, where the boundary between different classes of data is more easily identified. It is very computer intensive for classical computers to compute data in higher dimensions. However, by using quantum gates, the QSVM algorithm can easily perform these calculations exponentially faster than a classical SVM algorithm\textsuperscript{[7]}.

QSVM:
\begin{enumerate}
\item Apply a feature map which feeds the classical data into the quantum circuit. In the feature map, the gates encode the input data to the quantum circuit by applying rotations of an angle \(\theta\) .
\item Apply a quantum kernel function to the input data which transforms the data into a high-dimensional quantum feature space using different gates.
\end{enumerate}
\begin{figure}
    \centering
    \includegraphics[width=0.5\linewidth]{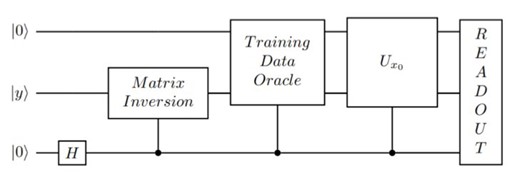}
    \caption{QSVM Circuit}
    \label{fig:2}
\end{figure}
Lastly, a Quantum Neural Net (QNN), unlike the other two, combines the principles of quantum computing and classical neural networks[9]. In a traditional neural network, artificial neurons are connected together in layers to process and analyze data. In a QNN, quantum gates are used to perform the same functions as the artificial neurons. The qubits are used to represent the input data, and the quantum gates are used to perform calculations on the data[9].

QNN:
\begin{enumerate}
    \item Apply a feature map that feeds the classical data into the quantum circuit. In the feature map, the gates encode the input data to the quantum circuit by applying rotations of an angle \(\theta\) .
    \item The encoded data is then processed by a quantum circuit, which is measured at the end to obtain a classical result. The result is fed to an optimization algorithm to improve its performance and output parameters into the quantum circuit to adjust the angles of the quantum gates, optimizing the network.
\end{enumerate}

\begin{figure}
    \centering
    \includegraphics[width=0.5\linewidth]{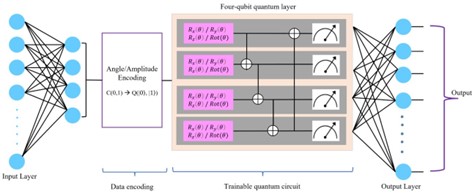}
    \caption{QNN Circuit}
    \label{fig:3}
\end{figure}

The main advantage of QNNs is that they can perform certain types of computations exponentially faster than classical neural networks. Again, this is because qubits can exist in a superposition of states, allowing them to represent multiple inputs at the same time, which can be processed in parallel. Additionally, the entanglement property of qubits allows for faster processing of the data\textsuperscript{[9]}.

 The invention of the qubit and the use of other quantum phenomena, such as quantum tunneling, superposition, and entanglement, have resulted in the development of quantum computers that are capable of carrying out some tasks far more quickly than classical computers\textsuperscript{[10]}. This has led to a great deal of interest in the field of quantum computing and its potential applications in fields such as cryptography, finance, drug discovery, and physics\textsuperscript{[1]}. 

 In this study, the application of interest is high energy physics, particularly particle collisions. Due to the nature of a particle collider, it can only detect the stable particles as all the unstable particles decay before they reach the detector. Many of the particles of interest are unstable, including the Higgs Boson. This paper focuses primarily on the Higgs Boson classifier problem and solving it with quantum classifiers.

\subsection{Purpose of Quantum Computers}
This paper explores the potential use of quantum computers in HEP. Quantum computers can revolutionize numerous fields. Quantum computers are capable of simulating intricate systems, such as chemical reactions, which could have significant implications for drug discovery and material science. Additionally, quantum computers could be utilized to optimize complicated processes, like supply chain management or financial portfolio optimization, producing more effective and profitable outcomes. 

\subsection{Gap, Focus, and Goal}
Research exists to show that quantum machine learning can perform comparably to the state-of-the-art machine learning methods that are currently used in particle physics[15,16,17]. The earliest studies applying QML on HEP state that due to the limitations in the datasets that classical machine learning methods are trained on, incorrect identification of background processes and other errors often occur[18]. To correct this, researchers have looked to quantum machine learning, as it has been theorized that quantum computers may outperform classical computers in machine learning tasks[19]. Furthermore, this has been proven by several quantum classifiers which achieved logarithmic speed up over classical algorithms[20].

However, the work done on QML in HEP has only resulted in equivalent performance compared to classical ML techniques thus far; findings have “demonstrated that the quantum and classical machine learning algorithms perform similarly”[17] and that “the quantum classifiers achieve results that are similar, if slightly better, than classical models trained on the same datasets”[16]. However, the potential for QML has been proven to be much higher than demonstrated by these studies[20]. This study aims to modify the algorithms used in previous studies to better fit this application. These modifications come from the recommendations of the previous studies and with these modifications, come potential performance that can outperform classical ML in HEP.

These modifications came from a previous study[17] since it promised the best performance compared to others on this topic. Several modifications are implemented into this study: the type of algorithms, feature map (which is how data is fed into the quantum circuit), number of qubits in the circuit, and the optimizer (which adjusts parameters on the circuit). All of these variables are recommended for future study in various papers[17].

By adding this research, we can prove QML may offer a ‘speed up’ advantage[21], which allows for quantum computing to replace classical computing in physics.

\section{Method}
\label{sec:main_section}
An evaluative research design was selected for this study because it involves the assessment of the effectiveness of modified classifiers on a specific dataset. This study aims to investigate whether modifications (to quantum classifiers) could lead to improved performance on the Higgs optimization problem. 

A systematic and objective evaluation of the modified classifiers was made possible by an evaluative research design. By comparing the improved classifiers’ performance with their traditional counterparts, one could determine whether the modifications result in a statistically significant performance improvement. 

Furthermore, the paired-sample t-tests utilized for data analysis enable one to determine whether the performance difference between the improved and non-improved quantum classifiers is statistically significant. This technique ensures that the observations are based on sound statistical principles and can be applied to a larger population.[25]

\subsection{Modifications}
The modifications that are implemented in the study are the type of algorithms, feature map, number of qubits in the circuit, and the optimizer.
The algorithms that are used in this study are the VQC, QSVM, and QNN. In order to implement these algorithms, the machine learning library, Sci-kit Learn [18], and the quantum computing library, Qiskit[22] will be used. 
There are two feature maps used in this study; the Z feature map, which is simple and does not offer quantum advantage, and the ZZ feature map. The Z feature map is what was used in previous studies and is used in the pre-modification algorithms. The ZZ feature map is more complex and will be used in the modified algorithms.
The number of qubits in each algorithm will be increased from the traditional 8 qubits to 30 qubits. 30 was chosen to match the dataset that was used in this study.
Two optimizers were used in this study, COBYLA or SPSA, with the former being used in previous studies as it is less complex. From the recommendations of Anton Dekusar[22], the SPSA will be used for the modified algorithms.

\subsection{Data Acquisition}

For the purpose of algorithm training, we will utilize the training dataset provided by ATLAS, a detector at CERN. This dataset is available on Kaggle (an ML challenges website) and CERN’s open data portal. This dataset is composed of 250,000 simulated events, each having 30 features describing each event.
It is worth pointing out that there are not many signal events in the LHC, with just around two signal events per every one thousand events. Because of this, to make sure the algorithm training is efficient, the simulation has been designed with more signal events (~125,000 events out of 250,000), which is accounted for by the weight column.
The signal event is the decay of Higgs Boson to tau-tau (another type of fundamental particle). This decay was chosen due to its low probability of occurring 0.1\%. (Figure 4 is a Reyman diagram of the decay process and Figure 5 is a graphic of the collision) The background events are any other collisions that are taking place. 

\begin{figure}
    \centering
    \includegraphics[width=0.5\linewidth]{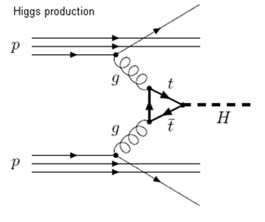}
    \caption{A Higgs Boson decaying to two tau[23}
    \label{fig:4}
\end{figure}
Thus, the utilization of this dataset provides a comprehensive range of features and labels that can be used to train and optimize machine learning algorithms. This simulation is a great dataset for this study, letting one evaluate the effectiveness of the algorithms in accurately classifying Higgs Bosons. 

\subsection{Data Pre-processing }
The data was split into three sections; 80\% will be used for the training set, 10\% will be the validation set and the other 10\% will be the test set. These section splits are commonly used in machine learning applications.
Principal Component Analysis (PCA) was used to lower the number of features in the data, which is done by identifying the most important features that contribute the most variance in the data. PCA analysis can be justified in this study as the dataset used in this study consists of 30 features, which would typically restrict the number of qubits in usage. Using a PCA analysis can enable one to reduce the number of features in the dataset, allowing one to vary the amount of qubits with the same dataset.

\subsection{Evaluating the Algorithms }
In order to evaluate the algorithms, three factors are evaluated: area under the ROC curve, F1-score, and the accuracy sci-kit learn[18] will be used to implement the performance metrics. These factors all use a scale of 0 to 1, with 1 being a perfect classifier, 0.5 being random guessing and 0 being the worst case. These metrics will allow for a comparison between the algorithms before and after these modifications.
Each classifier outputs the probability of each event being a Higgs Boson and a background event. A ROC (receiver operating characteristic) curve changes the confidence threshold to be considered as a Higgs Boson. The accuracy of these predictions given the various confidence thresholds is then plotted. The area under this curve, known as the AUC, then gives a measure of how accurate the classifier is across all confidence thresholds, meaning a high AUC indicates better classifier performance. AUC measures the tradeoff between the true positive rate and the false positive rate, providing an overall estimate of classifier performance.
The F1 score can be used to assess the performance of the classifier on both the signal and background events. The F1-score provides a balanced classifier evaluation by considering both precision and recall. 
The F1 score is defined as the \(F1=TP/(TP+(1/2)(FP+FN))\)
where TP represents the true positives, FP represents the false positives, and FN represents the false negatives.

\begin{table}
    \centering
    \begin{tabular}{|c|c|c|c|} \hline 
         &  &  \multicolumn{2}{|c|}{Actual}\\ \hline 
         &  &  Positive (Signal Events)& Negative (Background Events)\\ \hline 
         Predicted&  True (Signal Events)&  True Positive& True Negative\\ \hline 
         Predicted&  False (Background Events)&  False Positive& False Negative\\ \hline
    \end{tabular}
    \caption{\textit{Predicted values versus actual values} }
    \label{tab:1}
\end{table}
Lastly, the accuracy of each classifier will be determined. This is done by taking the number of correctly predicted signal events divided by the total number of signal events.

The algorithms were run without modifications and then run with modifications. Metrics were collected in both runs to allow for comparison. For this comparison, a t-test was used. A t-test compares two groups of data to see if one is bigger than the other. When the p-value gets closer to zero, group 2 is more likely to be bigger than group 1. In this study, group 1 refers to the performance of algorithms before modifications and group 2 refers to the performance after modifications.

\section{Results}
\label{sec:modelling}
\subsection{Comparisons of the Algorithms}
The evaluation was based on three performance metrics: accuracy, area under the curve (AUC), and F1-score. The results showed that the modifications made to the classifiers led to improvements in their performance on all three metrics.

\begin{table}
    \centering
    \begin{tabular}{|c|c|c|c|c|c|c|} \hline 
         &  \multicolumn{2}{|c|}{VQC}&  \multicolumn{2}{|c|}{QSVM}&  \multicolumn{2}{|c|}{QNN}\\ \hline 
         &  Before&  After&  Before&  After&  Before& After
\\ \hline 
         Accuracy
&  0.720558882&  0.74850299&  0.687083888&  0.82455089&  0.704590818& 0.708582834
\\ \hline 
         F1-Score
&  0.798437882&  0.81349085&  0.612978314&  0.89509695&  0.613609075& 0.634395884
\\ \hline 
         AUC
&  0.76&  0.82&  0.77&  0.85&  0.76& 0.78
\\ \hline
    \end{tabular}
    \caption{Table of all performance metrics before and after modifications}
    \label{tab:2}
\end{table}
The modified versions of the variational quantum circuit and quantum support vector machine outperformed their unmodified counterparts in terms of accuracy and AUC. Specifically, the modified variational quantum circuit’s accuracy increased by 3\% and the AUC increased 6\%. Similarly, the modified quantum support vector machine had an accuracy increase of 14\% and an AUC increase of 8\%. However, the quantum neural nets did not show a significant improvement, improving only by 0.4\% in accuracy and 2\% in AUC.
Before doing a t-test, one has to ensure that the conditions — independent and random — for the test to work are met. As per the central limit theorem, a sample size of at least 30 is necessary for the sample to be considered normal. However, in this scenario, resource limitations make obtaining a sample size of 30 different algorithms for each treatment infeasible. As a result, the reliability and accuracy of the t-test results may be affected, and this factor should be taken into account.

\begin{figure}
    \centering
    \includegraphics[width=0.75\linewidth]{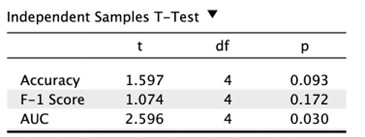}
    \caption{Result of the t-test}
    \label{fig:5}
\end{figure}

After conducting the t-test, one can determine if the improvement is significant or likely to have happened randomly. A thing to note is that the F1-score has the highest p-value, indicating that the f1-score provides weak evidence against the null hypothesis, which is the hypothesis that there is no difference before and after modifications.
The modified VQC and QSVM showed significant improvements in the F1-score compared to their unmodified counterparts, indicating that the modifications improved the ability of these classifiers to detect positives (Higgs Bosons) while maintaining high precision. The F1-score does not care about how many negative examples there are, which makes it a useful metric when working on problems with heavily imbalanced datasets and when one cares more about detecting positives than detecting negatives (problems like the Higgs Bosons). 
The modified versions of the variational quantum circuit and quantum support vector machine showed improvements in accuracy compared to their unmodified counterparts, indicating that the modifications improved the ability of these classifiers to correctly detect Higgs Bosons. However, one must look at accuracy and AUC combined as accuracy can be problematic, especially when the dataset is imbalanced. The dataset in this study was designed with an increased number of signal events, so it is not a concern in this study. However, in a real-life application, the data will be imbalanced, which should be taken into consideration when using the accuracy metric.
The improvement in AUC could be due to a variety of factors; nonetheless, an increase in AUC, F1-Score, and accuracy demonstrates that the classifiers are better at distinguishing between signal and background events.
After the modifications, both VQC and QSVM showed improvement, but the QSVM displayed a more substantial enhancement. Moreover, after the modifications, the QSVM had the highest scores in all performance metrics. It is possible that the modifications implemented were more effective for the QSVM than for the VQC, leading to a greater improvement in the former's performance. Alternatively, the inherent differences between the two classifiers may have influenced the improvement observed.
On the other hand, the QNN did not show significant improvement. The QNN’s metrics did increase slightly, however, the increase was relatively small. It is possible that the modifications did not address the specific weaknesses of the QNN or that the QNN may require more extensive modifications to demonstrate significant improvement in performance.
\subsection{Comparisons of Other Algorithms}
On average, the performance of these quantum algorithms was below that of the classical algorithms. However, the QSVM algorithm outperformed several other well-known classifiers. For example, it showed better results than the popular XGBoost algorithm (Chen and He) in terms of both accuracy and F1-score. The XGBoost algorithm is a classical algorithm that was among the top 2\% of algorithms in the 2014 Higgs ML Challenge.
It is also worth noting that the top accuracy for the best-performing algorithm was 84\% (Ahmad, 2015) while the QSVM accuracy was 82.4\%.
Despite its superior performance in terms of accuracy and F1-score, the QSVM algorithm's AUC was lower than that of the XGBoost algorithm as the XGBoost algorithm had an AUC of 0.904, whereas the best AUC achieved by the QSVM algorithm (after modifications) was only 0.85.
\subsection{Revisiting the Hypothesis}
With this data, the null hypothesis can be rejected as the modifications provided a statistically significant increase in the performance of algorithms to varying degrees.

\section{Conclusions and Future Directions}
\label{sec:discussion}
\subsection{Conclusions}
This study evaluated modifications to quantum classifiers to better identify Higgs Bosons using simulated data. The study explored three supervised classification algorithms and compared the modified performance with non-modified algorithms and other classical algorithms. It was found that the QSVM had the best performance increase after the modifications in terms of accuracy, AUC, and F1-Score. Although the QVC did not perform as well as the QSVM, it was still capable of performing at comparable accuracies and improved after modifications.

One of the modifications specified an increase in the number of qubits which may benefit both QSVM and QVC as they rely on the creation of high-dimensional feature vectors to improve their classification accuracy. More qubits in a quantum circuit can create a larger feature space, allowing for a more fine-grained representation of the input data, which in turn leads to improved classification.

 While QSVM is designed specifically for classification tasks and has been shown to outperform classical machine learning algorithms in some cases, QVC is a more general-purpose quantum circuit. Furthermore, QSVM has an advantage in handling non-linear data, which could benefit in the identification of Higgs Bosons. 

However, the QNN did not show significant improvement after modifications. In some cases, adding more qubits may introduce additional errors in the computation, making it more difficult to train the QNN effectively. Additionally, increasing the number of qubits may also increase the complexity of the quantum gates and circuits required to implement the QNN, making it harder to optimize the circuit.

These results demonstrate that the modifications proposed had some improvement in the ability to classify Higgs Bosons in a realistic HEP analysis. 

These modifications can lead to the future development of quantum machine learning for Higgs Boson identification, potentially leading to a replacement of classical classifiers with their quantum counterparts. This can aid in the verification of the Standard Model of particle physics and potentially the discovery of new physics beyond the Standard Model. 

Moreover, these results contribute insights into the development of quantum machine learning for particle physics applications beyond the Higgs Boson, as well as for other fields that employ similar machine learning techniques. 
\subsection{Limitations \& Future Directions }
The dataset that was used to train and test the models was very complex, but it is not a perfect simulation of a particle collision.  However, as it was simulated with CERN’s simulator rather than third party simulators, the simulation is more complex and more realistic. The dataset also only covers one type of particle decay (higgs to tau-tau), but Higgs Bosons can decay into other particles. Further studies can expand to include other Higgs Boson decays. Furthermore, studies can also expand to include datasets that are imbalanced as the dataset used in this study was balanced.
Additionally, as the study focused only on three algorithms, further research could explore and test other algorithms. These future studies can also explore different implementations of the algorithms; this study uses Qiskit’s basic libraries however future studies can use other libraries like Penny Lane or PegasusSVM.
The results also show that the improvement of quantum neural nets was restricted, thus a future study could focus on developing enhanced modifications for quantum neural nets. A suggestion for future studies looking at improving the QNN is to look at using Qiskit’s Sampler primitive rather than Qiskit’s Estimator. The sampler primitive just allows the user to have more manual control over the circuit whereas the estimator primitive does a lot of the calculations by itself.[22] 
Overall, this study yields statistically significant results and effectively fills the gap of fitting QML in the identification of Higgs Boson. It was shown that the QSVM classifier used outperforms current date technology, and future improvements can involve quantum classifier design. The content and research performed in this paper are a step to bringing future improvement into the quantum computing industry by improving both efficiency and accuracy.

\hfill

\textit{Acknowledgments} \\I would like to express my sincere appreciation to Dr. Christopher Hill, Professor of Physics at OSU, for his invaluable guidance throughout the study.\\

%\break
\section{Work Cited}
[1] C. Mariotti, “Observation of a new boson at a mass of 125 Gev with the CMS experiment at the LHC,” The Thirteenth Marcel Grossmann Meeting, 2015. 

[2] M. Gaillard, “CERN Data Centre passes the 200-Petabyte milestone,” CERN Document Server, 02-Aug-2017. [Online]. Available: https://cds.cern.ch/record/2276551. [Accessed: 05-Oct-2022].

[3] B. Laforge and ATLAS Collaboration, “Search for a low mass standard model higgs boson with the Atlas Detector at the LHC,” AIP Conference Proceedings, 2013. 

[4] P. Baldi, K. Cranmer, T. Faucett, P. Sadowski, and D. Whiteson, “Parameterized machine learning for high-energy physics,” arXiv.org, 28-Jan-2016. [Online]. Available: https://arxiv.org/abs/1601.07913. [Accessed: 23-Jan-2023].

[5] J. Preskill, “Quantum Computing in the NISQ era and beyond,” Quantum, vol. 2, p. 79, 2018. 

[6] M. Campbell-Kelly and W. Aspray, Computer: A History of the Information Machine, New York Basic Books, 1996, hardback ISBN 0-465-02989-2, 28, paperback ISBN 0-464-02990-6

[7] M. A. Nielsen and I. L. Chuang, Quantum Computation and Quantum Information. Cambridge: Cambridge University Press, 2022. 

[8] S. Aaronson and A. Arkhipov, “The computational complexity of linear optics,” ACM Digital Library, 2011. [Online]. Available: https://doi.org/10.1145/1993636.1993682. [Accessed: 25-Apr-2023]. 

[9] V. Havlicek, A. D. Córcoles, K. Temme, A. W. Harrow, A. Kandala, J. M. Chow, and J. M. Gambetta, “Supervised learning with quantum enhanced feature spaces,” arXiv.org, 05-Jun-2018. [Online]. Available: https://arxiv.org/abs/1804.11326. [Accessed: 23-Jan-2023].

[10] A. Gadotti, F. Houssiau, L. Rocher, B. Livshits, and Y.-A. de Montjoye, “When the signal is in the noise: Exploiting Diffix's sticky noise,” arXiv.org, 29-Oct-2019. [Online]. Available: https://arxiv.org/abs/1804.06752. [Accessed: 23-Jan-2023].

[11] M. Herschend, Y. Liu, and H. Nakaoka, “$N$-exangulated categories,” arXiv.org, 10-Dec-2018. [Online]. Available: https://arxiv.org/abs/1709.06689. [Accessed: 23-Jan-2023].

[12] A. K. Ekert, “Quantum cryptography and computation,” Advances in Quantum Phenomena, pp. 243–262, 1995.

[13] A. Mott, J. Job, J.-R. Vlimant, D. Lidar, and M. Spiropulu, “Solving a higgs optimization problem with quantum annealing for machine learning,” Nature, vol. 550, no. 7676, pp. 375–379, 2017. 

[14] V. Belis, S. González-Castillo, C. Reissel, S. Vallecorsa, E. F. Combarro, G. Dissertori, and F. Reiter, “Higgs analysis with Quantum Classifiers,” EPJ Web of Conferences, vol. 251, p. 03070, 2021. 

[15] S. L. Wu, J. Chan, W. Guan, S. Sun, A. Wang, C. Zhou, M. Livny, F. Carminati, A. Di Meglio, A. C. Li, J. Lykken, P. Spentzouris, S. Y.-C. Chen, S. Yoo, and T.-C. Wei, “Application of quantum machine learning using the quantum variational classifier method to high energy physics analysis at the LHC on IBM Quantum Computer Simulator and hardware with 10 qubits,” Journal of Physics G: Nuclear and Particle Physics, vol. 48, no. 12, p. 125003, 2021. 

[16]J. Biamonte, P. Wittek, N. Pancotti, P. Rebentrost, N. Wiebe, and S. Lloyd, “Quantum Machine Learning,” Nature, vol. 549, no. 7671, pp. 195–202, 2017. 

[17] Z. Abohashima, M. Elhosen, E. H. Houssein, and W. M. Mohamed, “Classification with Quantum Machine Learning: A Survey,” arXiv.org, 22-Jun-2020. [Online]. Available: https://arxiv.org/abs/2006.12270. [Accessed: 05-Oct-2022].

[20] F. Pedregosa, G. Varoquaux, A. Gramfort, V. Michel, B. Thirion, O. Grisel, M. Blondel, A. Müller, J. Nothman, G. Louppe, P. Prettenhofer, R. Weiss, V. Dubourg, J. Vanderplas, A. Passos, D. Cournapeau, M. Brucher, M. Perrot, and É. Duchesnay, “Scikit-Learn: Machine learning in Python,” arXiv.org, 05-Jun-2018. [Online]. Available: https://arxiv.org/abs/1201.0490. [Accessed: 25-Apr-2023]. 

[21] Q. Contributors, “Qiskit: An Open-source Framework for Quantum Computing,” GitHub, 2023. [Online]. Available: https://raw.githubusercontent.com/Qiskit/qiskit/master/Qiskit.bib. [Accessed: 2023]. 

[22] A. Dekusar, Qiskit Slack forum. (2023, January 22). How to implement a QNN \& QSVM in Qiskit? [Online forum]. Retrieved from https://qiskit.slack.com/archives/CB6C24TPB/p1676381806300299

[23] A. V. Gritsan, R. Röntsch, M. Schulze, and M. Xiao, “Constraining anomalous higgs boson couplings to the heavy-flavor fermions using matrix element techniques,” Physical Review D, vol. 94, no. 5, 2016. 

[24] R. M. Bianchi, “Higgs candidate decaying to 2 tau leptons in the Atlas Detector,” CERN Document Server, 26-Nov-2013. [Online]. Available: https://cds.cern.ch/record/1631395?ln=en. [Accessed: 25-Apr-2023]. 

[25] H. Abdi and L. J. Williams, “Principal component analysis,” Wiley Interdisciplinary Reviews: Computational Statistics, vol. 2, no. 4, pp. 433–459, 2010. 

[26] B. Gerald, “A brief review of independent, dependent and one sample t-test,” International Journal of Applied Mathematics and Theoretical Physics, vol. 4, no. 2, p. 50, 2018.

\end{document}